\documentclass[pra,amsmath,amsfonts]{revtex4-1}
\usepackage{epsf,epsfig}

\usepackage{graphicx}
\usepackage{dcolumn}
\usepackage{bm}

\usepackage{color,soul}
\usepackage{xcolor}	
\usepackage[hang,raggedright]{subfigure}

\newtheorem{theorem}{Theorem}
\newtheorem{lemma}[theorem]{Lemma}

\newenvironment{proof}[1][Proof]{\begin{trivlist}
\item[\hskip \labelsep {\bfseries #1}]}{\end{trivlist}}

\newcommand{\ket}[1]{|#1\rangle}
\newcommand{\bra}[1]{\langle #1|}
\newcommand{\project}[1]{\ket{#1}\bra{#1}}

\newcommand{\Id}{{\mathbb I}}

\newcommand{\Tr}{{\mathrm {Tr}}}

\newcommand{\etal}{\textit {et al.} }

\newcommand{\w}{\mathrm{w}}
\newcommand{\iso}{\mathrm{iso}}

\begin{document}


\title{Comparison of quantum discord and fully entangled fraction  of two classes of $d\otimes d^2$ states}

\author{Javad Behdani}
 \email{behdani.javad@stu.um.ac.ir}
\affiliation{Department of Physics, Ferdowsi University of Mashhad, Mashhad, Iran}
\author{Seyed Javad Akhtarshenas}
 \email{akhtarshenas@um.ac.ir}
\affiliation{Department of Physics, Ferdowsi University of Mashhad, Mashhad, Iran}
\author{Mohsen Sarbishaei}
 \email{sarbishei@um.ac.ir}
\affiliation{Department of Physics, Ferdowsi University of Mashhad, Mashhad, Iran}


\begin{abstract}
The quantumness of a generic state is the resource of many applications in quantum information theory and it is interesting to survey the measures which are able to detect its trace in the properties of the state. In this work we study the quantum discord and fully entangled fraction of two classes of bipartite states and compare their behaviors. These classes are complements to the $d\otimes d$ Werner and isotropic states, in the sense that each class possesses the same purification as the corresponding complemental class of states. Our results show that   maximally entangled mixed states are also maximally discordant states, leading to a generalization of the well-known fact that all maximally entangled pure states  have also maximum quantum discord. Moreover, it is shown that the separability-entanglement boundary of a Werner or isotropic  state is manifested as an  inflection point in the diagram of quantum discord of the corresponding complemental state.


\end{abstract}



\maketitle


\section{Introduction}

The exploitation of quantum states would result in the stronger ability in the processing of information with respect to the case of using classical systems \cite{Nielsen&Chuang:2000}. The origin of the speed-up in quantum information processing is the quantum correlation of the states which itself originates from the strange properties of quantum mechanics. One of these properties is the superposition principle and can be emanated in the entanglement \cite{Horodecki&etal:2009}. Entanglement plays a significant role in the theory of quantum information and  quantum computation and was believed to be the only proper nominate for the quantum correlation, because of the efficiency of quantum algorithms in analogy to the classical ones, when they are applied alongside with entangled states. However, entanglement is not able to represent all of the quantumness of correlations,  in the sense that  there are some disentangled states which can  be used in efficient quantum algorithms \cite{Knill&Laflamme:1998,Datta&etal:2005,Datta&Vidal:2007}. Therefore,  another strange property of quantum systems, i.e. the measurement, is considered which leads to a measure of quantum correlation called quantum discord \cite{Ollivier&Zurek:2001}. Aside the advantages and disadvantages of any of these two concepts, considering both of them will give us stronger vision about the quantumness of system.

Because of the importance of the maximally entangled states in quantum information  tasks, it may be interesting to study  a measure which is able to determine the degree that the state is close to a maximally entangled state.  Fully entangled fraction, as a measure for the degree of being close to the  maximal entangled states, is defined for  a state $\rho$ acting on $d\otimes d$ Hilbert space as follows \cite{Bennett&etal:1996}
\begin{equation}
\label{FEF}
\mathcal{F}(\rho)=\max_{U,V}\bra{\psi^{\max}}(U\otimes V)\rho(U^\dag\otimes V^\dag)\ket{\psi^{\max}}=\max_{V}\bra{\psi^{\max}}(\mathbb{I}\otimes V)\rho(\mathbb{I}\otimes V^\dag)\ket{\psi^{\max}},
\end{equation}
where $\ket{\psi^{\max}}=\frac{1}{\sqrt{d}}\sum_{j=1}^d|jj\rangle$ is the maximally entangled pure state of the $d\otimes d$ system, $U$ and $V$ are unitary operators acting on the marginal subsystems and $\mathbb{I}$ denotes the identity operator acting on the first subsystem. Although all of the maximally entangled states of the  $d\otimes d$ bipartite systems are pure, for a general $d\otimes d^\prime$ system there are mixed states which are maximally entangled. Let $\rho$ be an arbitrary state, acting on $d\otimes d^\prime$  Hilbert space  with $d\le d^\prime$, and define $K$ via  $d^\prime=Kd+r$ where $0\leq r<d$ and $K\geq 1$. A generalization of Eq. \eqref{FEF} is then proposed   as the maximum overlap of the state $\rho$ with $K$ maximally entangled pure states as follows \cite{Zhao:2015}
\begin{equation}
\label{FEFG}
\mathcal{F}(\rho)=\max_{U,V}\sum_{i=1}^K\bra{\psi^{\max}_i}(U\otimes V)\rho(U^\dag\otimes V^\dag)\ket{\psi^{\max}_i}=\max_{V}\sum_{i=1}^K\bra{\psi^{\max}_i}(\mathbb{I}\otimes V)\rho(\mathbb{I}\otimes V^\dag)\ket{\psi^{\max}_i},
\end{equation}
where, in the last term, maximum is taken over all unitary  operators acting on the  second subsystem. Moreover, $\{\ket{\psi^{\max}_i}\}_{i=1}^{K}$ is the set of maximally entangled pure states of the $d\otimes d^\prime$ system, defined by
\begin{equation}
\ket{\psi^{\max}_i}=\frac{1}{\sqrt{d}}\sum_{j=1}^d\ket{j}\otimes\ket{j+(i-1)d}.
\end{equation}
It is shown that $\frac{K}{dd^\prime}\leq\mathcal{F}(\rho)\leq1$ for all $d\otimes d^\prime$ quantum states, while $\frac{K}{dd^\prime}\leq\mathcal{F}(\rho)\leq\frac{1}{d}$, for all $d\otimes d^\prime$ separable states \cite{Zhao:2015}. Although fully entangled fraction cannot be regarded as a measure of entanglement in its own right, it may be used  as a quantifier of entanglement applications. Historically, the first such usage  was worked out by Horodecki \etal \cite{Horodecki&etal:1999}, where the authors found a  relationship between fully entangled fraction and the optimal fidelity of teleportation.  In addition, in a two-qubit system, the fully entangled fraction can also be related to the fidelities of other important quantum information tasks such as dense coding, entanglement swapping and quantum cryptography in such a way as to provide an inclusive measure of these entanglement applications \cite{Grondalski&etal:2002}. Additionally, it can be exploited as an index to characterize the non-local correlation \cite{Zhou&Guo:2000}.

Quantum discord, introduced  by Ollivier and Zurk in their seminal work  \cite{Ollivier&Zurek:2001}, is defined for a generic bipartite state $\rho^{AC}$ as follows
\begin{equation}
\mathcal{D}^{\leftarrow}(\rho^{AC})=I(\rho^{AC})-\mathcal{I}^{\leftarrow}(\rho^{AC}),
\label{QD}
\end{equation}
where $I(\rho^{AC})=S(\rho^A)+S(\rho^C)-S(\rho^{AC})$ is the mutual information of $\rho^{AC}$, and
\begin{equation}
\mathcal{I}^{\leftarrow}(\rho^{AC})=S(\rho^A)-\sum_{i}p_iS(\rho^{A|\{\Pi_i^C\}}),
\end{equation}
is the classical correlation between parts of $\rho^{AC}$.
Here $\{\Pi_i^C\}=\left\{\project{i}\right\}$ is the set of projection operators on the subsystem $C$, $p_i=\Tr[(\mathbb{I}\otimes\Pi_i^C)\rho^{AC}(\mathbb{I}\otimes\Pi_i^C)]$  is the probability of the $i$-th outcome for the subsystem $C$, and $\rho^{A|\{\Pi_i^C\}}=\frac{1}{p_i}\Tr_C[(\mathbb{I}\otimes\Pi_i^C)\rho^{AC}(\mathbb{I}\otimes\Pi_i^C)]$ represents the post-measurement state. An independent work by Henderson and Vedral leads to a similar measure \cite{Henderson&Vedral:2001}. Finding the quantum discord of an arbitrary state is still an open problem in quantum information theory and the analytical formula for the measure has been found just for a limited class of states \cite{Henderson&Vedral:2001,Ollivier&Zurek:2001,Luo:2008,Rulli&Sarandy:2011,Okrasa&Walczak:2011,Xu:2013}.

Let $|\psi^{ABC}\rangle$ be a pure state of the Hilbert space $\mathcal{H}^{A}\otimes\mathcal{H}^{B}\otimes\mathcal{H}^{C}$. It is shown that the following relation can be found between entanglement of formation \cite{Bennett&etal:1996} of the bipartite state $\rho^{AB}=\Tr_{C}{|\psi^{ABC}\rangle\langle\psi^{ABC}|}$  and the so-called post-measurement mutual information of the bipartite state $\rho^{AC}=\Tr_{B}{|\psi^{ABC}\rangle\langle\psi^{ABC}|}$  \cite{Koashi&Winter:2004}
\begin{equation}
E_f(\rho^{AB})+\mathcal{I}^{\leftarrow}(\rho^{AC})=S(\rho^{A}),
\label{KWR}
\end{equation}
where $E_f(\cdot)$ is the entanglement of formation of the state and $\mathcal{I}^{\leftarrow}(\cdot)$ is the right post-measurement mutual information. The  state $\rho^{AB}$ is called the $C$ complement to $\rho^{AC}$, and similarly, $\rho^{AC}$ is called the $B$ complement to $\rho^{AB}$. From  Eqs. \eqref{QD} and \eqref{KWR} one  infer that
\begin{eqnarray}\nonumber
\mathcal{D}^{\leftarrow}(\rho^{AC})&=&E_f(\rho^{AB})+S(\rho^C)-S(\rho^{AC}) \\
&=& E_f(\rho^{AB})+S(\rho^{AB})-S(\rho^{B}),
\label{QDKW}
\end{eqnarray}
where the second line follows from $S(\rho^{C})=S(\rho^{AB})$ and $S(\rho^{AC})=S(\rho^{B})$.
This equation leads to an alternative way for calculating the quantum discord of a class of  states, without the hard process of optimization which is a part of the definition of quantum discord. Evidently,  the use of this method is limited to the states whose entanglement of formation of their complemental states is known. However, the strength of the mentioned method was shown in Ref. \cite{Shi&etal:2011} where the authors could find the quantum discord of some two-qubit rank-2 states.

The aim of this paper is to compare the fully entangled fraction and quantum discord of two classes of $d\otimes d^2$ states. The states considered here are complement to the $d\otimes d$ Werner and isotropic states.  Using  Eq. \eqref{QDKW} we  obtain a closed relation for the  quantum discord of these states. We also use the notion of fully entangled fraction and find an exact relation for the first class of our $d\otimes d^2$ states, i.e. states complement to the Werner states. For the second  class of $d\otimes d^2$ states, i.e. states complement to $d\otimes d$ isotropic states, we provide a lower bound for the fully entangled fraction. A comparison of the measures  shows that the quantum discord and fully entangled fraction of each class behave similarly. In particular, our results show that maximally entangled mixed states are also maximally discordant states. This generalizes the well known fact that all maximally entangled pure states  have also maximum quantum discord. In addition, our results show that  the separability-entanglement boundary of a Werner or isotropic  state is manifested as an  inflection point in the diagram of quantum discord of the corresponding complemental state. Therefore, further insight into the notion of separability-entanglement paradigm may be provided by investigating  quantum discord of the complemental state.

The remainder of the paper is arranged as follows. In  Section \ref{s2}, we  present a class of $d\otimes d^2$ states complement to  $d\otimes d$ Werner states and provide an exact solution for their fully entangled fraction and quantum discord. Section \ref{s3} is devoted to  the $d\otimes d^2$ states complement to $d\otimes d$ isotropic states, and finding a lower bound for fully entangled fraction. In Section  \ref{s4} we provide a discussion by presenting some examples, i.e. the cases $d=2,3$,  and compare the fully entangled fraction and quantum discord of the mentioned states. We conclude the paper in Section \ref{s5}.

\section{The $d\otimes d^2$ states complement to the $d\otimes d$ Werner states}
\label{s2}
A generic  $d\otimes d$ Werner state acting on $\mathcal{H}^A\otimes\mathcal{H}^B$ is defined by
\begin{equation}
\rho_\w^{AB}=\frac{d-x}{d^3-d}\mathbb{I}_{d^2}+\frac{dx-1}{d^3-d}F,\quad x\in[-1,1],
\label{mmwerner}
\end{equation}
where $F=\sum_{k,l=1}^d|kl\rangle\langle lk|$, and $x=\Tr{(\rho_\w^{AB} F)}$. It is shown that Werner state is separable if and only if $0\le x\le 1$ \cite{Vollbrecht&Werner:2001}. The Werner state \eqref{mmwerner} has the following spectral decomposition
\begin{equation}
\label{pmmwerner}
\rho_\w^{AB}=\lambda_{+}P^{(+)}+\lambda_{-}P^{(-)}.
\end{equation}
Here $\lambda_{\pm}=\frac{1\pm x}{d(d\pm 1)}$, and  $P^{(+)}$ and $P^{(-)}$ are projection operators on the $d_{+}=d(d+1)/2$- and $d_{-}=d(d-1)/2$-dimensional symmetric and antisymmetric subspaces of $\mathcal{H}^A\otimes\mathcal{H}^B$, respectively, defined by
\begin{equation}
P^{(+)}=\sum_{k\le l}^d\ket{\lambda_{kl}^{(+)}}\bra{\lambda_{kl}^{(+)}},\qquad P^{(-)}=\sum_{k<l}^d\ket{\lambda_{kl}^{(-)}}\bra{\lambda_{kl}^{(-)}},
\end{equation}
where
\begin{equation}
\ket{\lambda_{k\le l}^{(+)}}=N_{kl}(\ket{kl}+\ket{lk}), \qquad \ket{\lambda_{k<l}^{(-)}}=N_{kl}(\ket{kl}-\ket{lk}),
\end{equation}
with $N_{kk}=\frac{1}{2}$ and $N_{kl}=\frac{1}{\sqrt{2}}$ for $k\neq l$.
It follows that one can write the following purification for the $d\otimes d$ Werner state
\begin{eqnarray}
\ket{\Psi^{ABC}}=\sqrt{\lambda_{+}}\sum_{k\le l}^d\ket{\lambda_{kl}^{(+)}}\ket{\mu_{kl}^{(+)}}+\sqrt{\lambda_{-}}\sum_{k< l}^d\ket{\lambda_{kl}^{(-)}}\ket{\mu_{kl}^{(-)}},
\end{eqnarray}
where $\{\ket{\mu_{kl}^{(+)}}\}_{k\le l}$ and $\{\ket{\mu_{kl}^{(-)}}\}_{k < l}$  form an orthonormal basis for the $d^2$-dimensional Hilbert space $\mathcal{H}^{C}$.
 Accordingly, the $B$ complement to  the $d\otimes d$ Werner state can be written as follows
\begin{eqnarray}
\label{rhoacw}
\rho^{AC}=\frac{1}{d}\sum_{r=1}^{d}\ket{\Psi_r}\bra{\Psi_r},
\end{eqnarray}
with
\begin{eqnarray} \nonumber
\ket{\Psi_r}=\cos{\alpha}\ket{r}\ket{\mu_{rr}}
&+&\frac{\sin{\alpha}}{\sqrt{d-1}}\sum_{k=1}^{r-1}\ket{k}\left(\cos{\theta}\ket{\mu_{kr}^{(+)}}+\sin{\theta}\ket{\mu_{kr}^{(-)}}\right) \\
&+&\frac{\sin{\alpha}}{\sqrt{d-1}}\sum_{k=r+1}^{d}\ket{k}\left(\cos{\theta}\ket{\mu_{rk}^{(+)}}-\sin{\theta}\ket{\mu_{rk}^{(-)}}\right),\label{psir}
\end{eqnarray}
where $\cos{\alpha}=\sqrt{d\lambda_{+}}$, and $\cos{\theta}=\sqrt{\frac{\lambda_{+}}{\lambda_{+}+\lambda_{-}}}$.
Obviously,  the rank of the  state $\rho^{AC}$, acting on the $d\otimes d^2$-dimensional Hilbert space $\mathcal{H}^A\otimes \mathcal{H}^C$, equals to $d$. This follows easily from the fact that the $d$-dimensional Hilbert space $\mathcal{H}^B$ is needed to obtain the purified state $\ket{\Psi^{ABC}}$. For further use, let us present  the following  ordering for the  basis of the Hilbert space $\mathcal{H}^C$
\begin{eqnarray}\nonumber
& \{\ket{1}=\ket{\mu_{11}^{(+)}},\cdots,\ket{d}=\ket{\mu_{dd}^{(+)}}\}; & \\  \nonumber
& \{\ket{d+1}=\ket{\mu_{12}^{(+)}}, \cdots,\ket{d(d+1)/2}=\ket{\mu_{d(d-1)/2-1,d(d-1)/2}^{(+)}}\}; &\\ \label{HCBasis}
& \{\ket{d(d+1)/2+1}=\ket{\mu_{12}^{(-)}}, \cdots,\ket{d^2}=\ket{\mu_{d(d-1)/2-1,d(d-1)/2}^{(-)}}\}. &
\end{eqnarray}

\subsection{Fully entangled fraction}
Now in order to calculate fully entangled fraction, we can use Eq. \eqref{FEFG}. Looking at eigenvectors of $\rho^{AC}$, it is clear that all
vectors are written in their  Schmidt decomposition, with the same Schmidt rank $d$ and Schmidt coefficients  given by $\cos{\alpha}$ and $\frac{\sin{\alpha}}{\sqrt{d-1}}$, with  multiplicities  $1$ and $(d-1)$, respectively. Evidently, these states become maximally entangled states whenever  $x=1/d$, i.e. when   $\lambda_{+}=\lambda_{-}$ or equivalently  $\cos{\alpha}=\frac{1}{\sqrt{d}}$ and $\cos{\theta}=\frac{1}{\sqrt{2}}$, so that
\begin{eqnarray}\label{PsiMaxW}
\ket{\Psi_r^{\max}}=\frac{1}{\sqrt{d}}\sum_{k}^{d}\ket{k}\ket{\mu_k^{(r)}}, \quad \mathrm{for}\;\; r=1,\cdots,d,
\end{eqnarray}
where we have defined $\{\ket{\mu_k^{(r)}}\}_{k=1}^{d}$ (for $r=1,\cdots,d$)
\begin{eqnarray}
\ket{\mu_{r}^{(r)}}=\ket{\mu_{rr}^{(+)}}, \quad \ket{\mu_{k}^{(r)}}&=&\frac{1}{\sqrt{2}}\left(\ket{\mu_{kr}^{(+)}}+\ket{\mu_{kr}^{(-)}}\right)\;\; \mathrm{for}\;\; k<r, \\
 \ket{\mu_{k}^{(r)}}&=&\frac{1}{\sqrt{2}}\left(\ket{\mu_{rk}^{(+)}}-\ket{\mu_{rk}^{(-)}}\right) \;\; \mathrm{for}\;\; k>r,
\end{eqnarray}
as the local Schmidt basis of the vector $\ket{\Psi_r^{\max}}$ with respect to the part $C$.
On the other hand, as it is clear from the above equation, one can easily see that all of the above vectors  are pairwise orthogonal, i.e.
$\bra{\mu^{(r)}_{k}}\mu^{(s)}_l\rangle=\delta_{kl}\delta_{rs}$, so that they form an orthonormal basis for the $d^2$-dimensional Hilbert space $\mathcal{H}^C$.
\begin{lemma}\label{lem1}
For the $d\otimes d^2$ state $\rho^{AC}$ given in \eqref{rhoacw}, the set of  maximally entangled states $\{\ket{\chi_{k}}\}_{k=1}^{d}$ that maximize Eq. \eqref{FEFG} is given by the set  $\{\ket{\Psi_r^{\max}}\}_{r=1}^{d}$.
\end{lemma}
A proof for the above lemma is provided in the Appendix \ref{apa}. Accordingly, one can easily calculate $\mathcal{F}(\rho^{AC})$ as
\begin{eqnarray}\nonumber
\mathcal{F}(\rho^{AC})&=&\max_{V}\sum_{i=1}^{K}\bra{\psi_i^{\max}}(\Id\otimes V)\rho^{AC}(\Id\otimes V^\dagger)\ket{\psi_i^{\max}} \\
&=&\sum_{r=1}^{d}\bra{\Psi_r^{\max}}\rho^{AC} \ket{\Psi_r^{\max}}=\frac{1}{d}\sum_{r=1}^{d}|\bra{\Psi_r^{\max}}\Psi_r\rangle|^2=|\bra{\Psi_r^{\max}}\Psi_r\rangle|^2,
\end{eqnarray}
which after some calculations takes the following form
\begin{eqnarray}\label{FRhoACW}
\mathcal{F}(\rho^{AC})&=\frac{1}{4}\left((d+1)\sqrt{\lambda_{+}}+(d-1)\sqrt{\lambda_{-}}\right)^2=\frac{1}{2d}\left(d+x+\sqrt{(d^2-1)(1-x^2)}\right).
\end{eqnarray}

\subsection{Quantum discord}
For the state  $\rho^{AC}$  given by \eqref{rhoacw}, i.e. $B$ complement to  the $d\otimes d$ Werner state, we are able to find an exact expression for the quantum discord. Using
\begin{equation}
S(\rho_\w^{AB})=-\frac{d(d+1)}{2}\lambda_{+}\log_2{\lambda_{+}}-\frac{d(d-1)}{2}\lambda_{-}\log_2{\lambda_{-}},
\end{equation}
and $S(\rho^B)=\log_2{d}$ in Eq. \eqref{QDKW},  we find
\begin{eqnarray}\label{DRhoACW}
\mathcal{D}^{\leftarrow}(\rho^{AC})=E_f(\rho_\w^{AB})-\frac{1+x}{2}\log_2(\frac{1+x}{d^2+d})-\frac{1-x}{2}\log_2(\frac{1-x}{d^2-d})-\log_2{d}.
\end{eqnarray}
Here   $E_f(\rho_\w^{AB})$ is  entanglement of formation of the Werner states given by  \cite{Wootters:2001,Vollbrecht&Werner:2001}
\begin{equation}\label{EoFmmWerner}
E_f(\rho_\w^{AB})=\left\{\begin{array}{lll}h\left(\frac{1}{2}[1+\sqrt{1-x^2}]\right)&,&x\in[-1,0]\\0&,&x\in[0,1]\end{array}\right.,
\end{equation}
where $h(\cdot)$ is the Shannon binary  entropy defined by $h(x)=-x\log{x}-(1-x)\log{(1-x)}$.

A comparison of Eqs. \eqref{FRhoACW} and \eqref{DRhoACW} reveals that both quantities reach their maximal possible values $\mathcal{F}(\rho^{AC})=1$ and $\mathcal{D}(\rho^{AC})=\log_2 d$, respectively, only at the same point  $x_{\max}=\frac{1}{d}$, so that  maximally entangled mixed state is also maximally discordant state.

\section{The $d\otimes d^2$ states complement to  the $d\otimes d$ isotropic states}
\label{s3}
A $d\otimes d$ isotropic state acting on $\mathcal{H}^A\otimes \mathcal{H}^B$ is defined  by
\begin{equation}
\rho_\iso^{AB}=\frac{1-f}{d^2-1}\mathbb{I}_{d^2}+\frac{d^2f-1}{d^2-1}|\phi\rangle\langle\phi|,\quad f\in[0,1],
\end{equation}
where $|\phi\rangle=\frac{1}{\sqrt{d}}\sum_{k=1}^d|kk\rangle$, and $f=\bra{\phi}\rho_\iso^{AB}\ket{\phi}$. It is  known that $\rho_\iso^{AB}$ is separable for $f\leq \frac{1}{d}$ \cite{Terhal&Vollbrecht:2000}. The spectral decomposition of an isotropic state is given by
\begin{equation}
\rho_\iso^{AB}=\lambda_{\phi}P_{\phi}+\lambda_{\phi^\perp}P_{\phi^\perp},
\end{equation}
where $\lambda_{\phi}=f$ and $\lambda_{\phi^\perp}=\frac{1-f}{d^2-1}$. In addition,  $P_{\phi}$ and $P_{\phi^\perp}$ are projection operators on the one- and  $(d^2-1)$-dimensional subspaces of $\mathcal{H}^A\otimes\mathcal{H}^B$, respectively,  defined by
\begin{equation}
P_{\phi}=\ket{\lambda_{11}}\bra{\lambda_{11}},\qquad P_{\phi^\perp}=\Id_{d^2}-P_{\phi}=\sum_{k=2}^d\ket{\lambda_{kk}}\bra{\lambda_{kk}}+\sum_{k\ne l}^d\ket{\lambda_{kl}}\bra{\lambda_{kl}}
\end{equation}
with $\ket{\lambda_{11}}=\ket{\phi}=\frac{1}{\sqrt{d}}\sum_{k=1}^d\ket{kk}$, $\ket{\lambda_{kk}}=\frac{1}{\sqrt{2}}(\ket{kk}-\ket{11})$ for $k\neq 1$, and $\ket{\lambda_{kl}}=\ket{kl}$  for $k\neq l$.
A purification of the isotropic state is
\begin{eqnarray}
\label{PsiABCiso}
\ket{\Phi^{ABC}}=\sqrt{\lambda_{\phi}}\ket{\lambda_{11}}\ket{\mu_{11}}
+\sqrt{\lambda_{\phi_\perp}}\left(\sum_{k=2}^d\ket{\lambda_{kk}}\ket{\mu_{kk}}+\sum_{k\ne l}^d\ket{\lambda_{kl}}\ket{\mu_{kl}}\right),
\end{eqnarray}
where $\{\ket{\mu_{kl}}\}_{k,l=1}^{d}$ forms an orthonormal basis for the $d^2$-dimensional Hilbert space $\mathcal{H}^{C}$.
The B complement to the $d\otimes d$ isotropic  state is obtained as
\begin{equation}\label{RhoIsoAC}
\varrho^{AC}=p\ket{\Phi_1}\bra{\Phi_1}
+\frac{1-p}{d-1}\sum_{r=2}^{d}\ket{\Phi_r}\bra{\Phi_r}, \qquad p=\left(\frac{\lambda_{\phi}}{d}+\frac{3}{2}(d-1)\lambda_{\phi_\perp}\right),
\end{equation}
where
\begin{equation}\label{Phi1}
\ket{\Phi_1}=\cos{\beta}\ket{1}\left[\cos{\vartheta}\ket{\mu_{11}}-\frac{\sin{\vartheta}}{\sqrt{d-1}}\sum_{k=2}^{d}\ket{\mu_{kk}}\right]
+\frac{\sin{\beta}}{\sqrt{d-1}}\sum_{k=2}^{d}\ket{k}\ket{\mu_{k1}},
\end{equation}
and
\begin{equation}
\label{Phir}
\ket{\Phi_{r}}=\cos{\tilde{\beta}}\ket{r}\left[\cos{\tilde{\vartheta}}\ket{\mu_{11}}+\sin{\tilde{\vartheta}}\ket{\mu_{rr}}\right]
+\frac{\sin{\tilde{\beta}}}{\sqrt{d-1}}\sum_{k\ne r}^{d}\ket{k}\ket{\mu_{kr}}, \quad 2\le r\le d.
\end{equation}
Here
\begin{eqnarray}
\cos{\beta}&=&\sqrt{\frac{2\lambda_{\phi}+d(d-1)\lambda_{\phi_{\perp}}}{2\lambda_{\phi}+3d(d-1)\lambda_{\phi_{\perp}}}}, \qquad \cos{\vartheta}=\sqrt{\frac{2\lambda_{\phi}}{2\lambda_{\phi}+d(d-1)\lambda_{\phi_{\perp}}}},\label{betatheta} \\
\cos{\tilde{\beta}}&=&\sqrt{\frac{2\lambda_{\phi}+d\lambda_{\phi_{\perp}}}{2\lambda_{\phi}+d(2d-1)\lambda_{\phi_{\perp}}}}, \qquad \cos{\tilde{\vartheta}}=\sqrt{\frac{2\lambda_{\phi}}{2\lambda_{\phi}+d\lambda_{\phi_{\perp}}}}.\label{betathetatilde}
\end{eqnarray}

\subsection{Fully entangled fraction}
It is clear from Eqs. \eqref{Phi1} and \eqref{Phir} that all eigenvectors of $\varrho^{AC}$ are written in their  Schmidt decomposition, with the same Schmidt rank $d$. However, in contrary to the case of Werner state, the Schmidt coefficients of all $\{\ket{\Phi_r}\}_{r=1}^{d}$ are not equal. More precisely, the  Schmidt coefficients of $\ket{\Phi_1}$ are  given by $\cos{\beta}$ and $\frac{\sin{\beta}}{\sqrt{m-1}}$, with  multiplicities  $1$ and $(d-1)$, respectively, and are  given for $\ket{\Phi_{r\ge 2}}$ by $\cos{\tilde{\beta}}$ and $\frac{\sin{\tilde{\beta}}}{\sqrt{d-1}}$, with  multiplicities  $1$ and $(d-1)$, respectively. Moreover, the region that these states become maximally entangled states is not the same for $r=1$ and $r\ne 1$. Indeed, $\ket{\Phi_1}$ becomes maximally entangled state  whenever  $f=\frac{3-d}{d^2-d+2}$,  i.e. when   $\lambda_{\phi}=(3-d)\lambda_{\phi_{\perp}}$, which has physical solution only for $d=2,3$. In these cases we have
\begin{eqnarray}\label{Phi1Maxm2}
\ket{\Phi_1^{\max}}&=&\frac{1}{\sqrt{2}}\ket{1}\left[\frac{1}{\sqrt{2}}\ket{\mu_{11}}-\frac{1}{\sqrt{2}}\ket{\mu_{22}}\right]
+\frac{1}{\sqrt{2}}\ket{2}\ket{\mu_{21}},\quad \mathrm{for} \quad d=2, \\ \label{Phi1Maxm3}
\ket{\Phi_1^{\max}}&=&\frac{1}{\sqrt{3}}\ket{1}\left[-\frac{1}{\sqrt{2}}\ket{\mu_{22}}-\frac{1}{\sqrt{2}}\ket{\mu_{33}}\right]
+\frac{1}{\sqrt{3}}\ket{2}\ket{\mu_{21}}+\frac{1}{\sqrt{3}}\ket{3}\ket{\mu_{31}} \quad d=3,
\end{eqnarray}
On the other hand, $\{\ket{\Phi_r}\}_{r=2}^{d}$ become maximally entangled states whenever  $f=\frac{d}{2(d^2-1)+d}$,  i.e. when
$\lambda_{\phi}=\frac{d}{2}\lambda_{\phi_{\perp}}$, which has physical solution for any $d$, so that
\begin{eqnarray}\label{PhirMaxm}
\ket{\Phi_{r}^{\max}}&=&\frac{1}{\sqrt{d}}\ket{r}\left[\frac{1}{\sqrt{2}}\ket{\mu_{11}}+\frac{1}{\sqrt{2}}\ket{\mu_{rr}}\right]
+\frac{1}{\sqrt{d}}\sum_{k\ne r}^{d}\ket{k}\ket{\mu_{kr}}, \quad 2\le r\le d,
\end{eqnarray}
for an arbitrary $d$. As it is clear from Eq. \eqref{Phi1Maxm2}-\eqref{PhirMaxm}, the local Schmidt basis of the part $C$ lacks the essential property of being pairwise orthogonal, except for the case $d=2$, so that these states  cannot be used to optimize relation \eqref{FEFG}.  However, let us define the following set of maximally entangled states which possesses  the required property of being pairwise orthogonal with respect to the Schmidt basis of part $C$
\begin{eqnarray}
\ket{\eta_{r}^{\max}}&=&\frac{1}{\sqrt{d}}\sum_{k=1}^{d}\ket{k}\ket{\mu_{kr}}, \quad 1\le r\le d.
\end{eqnarray}
Using this set, one can  obtain a lower bound for the fully entangled fraction as
\begin{eqnarray}\nonumber
\mathcal{F}(\varrho^{AC})&=&\max_{V}\sum_{i=1}^{K}\bra{\psi_i^{\max}}(\Id\otimes V)\varrho^{AC}(\Id\otimes V^\dagger)\ket{\psi_i^{\max}} \\ \nonumber
&\ge&\sum_{r=1}^{d}\bra{\eta_r^{\max}}\varrho^{AC} \ket{\eta_r^{\max}}=p|\bra{\eta_1^{\max}}\Phi_1\rangle|^2
+\frac{1-p}{d-1}\sum_{r=2}^{d}|\bra{\eta_r^{\max}}\Phi_r\rangle|^2 \\
&=& p|\bra{\eta_1^{\max}}\Phi_1\rangle|^2
+(1-p)|\bra{\eta_{r\ne 1}^{\max}}\Phi_{r\ne 1}\rangle|^2,
\end{eqnarray}
which takes the following form
\begin{eqnarray}
\mathcal{F}(\varrho^{AC})\geq\frac{1}{d}\left(p\left(\cos{\beta}\cos{\vartheta}+\sqrt{d-1}\sin{\beta}\right)^2
+(1-p)\left(\cos{\tilde{\beta}}\sin{\tilde{\vartheta}}+\sqrt{d-1}\sin{\tilde{\beta}}\right)^2\right)
\end{eqnarray}

\subsection{Quantum discord}
Using  Eq. \eqref{QDKW}, we can  provide an exact expression for the quantum discord of the $d\otimes d^2$ state $\varrho^{AC}$ given by Eq. \eqref{RhoIsoAC}, i.e. $B$ complement to the  isotropic state,  as
\begin{eqnarray}
\mathcal{D}^{\leftarrow}(\varrho^{AC})=E_f(\rho_\iso^{AB})-f\log_2{f}-(1-f)\log_2{\left(\frac{1-f}{d^2-1}\right)}-\log_2{d},
\end{eqnarray}
where we have used
\begin{equation}
S(\rho_{\iso}^{AB})=-f\log_2(f)-(1-f)\log_2(\frac{1-f}{d^2-1}),
\end{equation}
and $S(\rho^B)=\log_2{d}$.
Here $E_f(\rho_\iso^{AB})$,  entanglement of formation of the isotropic state, is zero for $f\le 1/d$ and takes    \cite{Terhal&Vollbrecht:2000,Wootters:2001}
\begin{equation}
E_f(\rho_\iso^{AB})=\textrm{co}\big(h(\xi)+(1-\xi)\log_2(d-1)\big),
\end{equation}
for $f\ge 1/d$, where
\begin{equation}
\xi=\frac{1}{d}\left(\sqrt{f}+\sqrt{(d-1)(1-f)}\right)^2,
\end{equation}
and $\textrm{co}(x)$ shows any function of $x$ indicates the convex hull \cite{Terhal&Vollbrecht:2000,Wootters:2001}. It is shown that for $d=2$ we have
\begin{equation}
E_f(\rho_\iso^{AB})=\mathcal{E}(2f-1),
\end{equation}
where $\mathcal{E}(g)=h(\frac{1+\sqrt{1-g^2}}{2})$, while for $d=3$, the entanglement of formation is given by  \cite{Terhal&Vollbrecht:2000}
\begin{equation}
E_f(\rho_\iso^{AB})=\left\{\begin{array}{lll}0&,&x\leq\frac{1}{d} \\
h(\xi)+(1-\xi)\log_2(d-1)&,&\frac{1}{d}<x<\frac{4}{d^2}(d-1)\\
\frac{(x-1)d}{d-2}\log_2(d-1)+\log_2d&,&\frac{4}{d^2}(d-1)\leq x\leq 1\end{array}\right..
\end{equation}
In addition, the authors of Ref. \cite{Terhal&Vollbrecht:2000} conjectured that the same equation is valid  for other  values of $d$.

\section{Discussion}
\label{s4}
In this section we provide a comparison between  fully entangled fraction and quantum discord of the $d\otimes d^2$ states. To make our considerations more clear, let us start with the simple case $d=2$.

\begin{figure}
\centering
\subfigure[]{\includegraphics[scale=0.7]{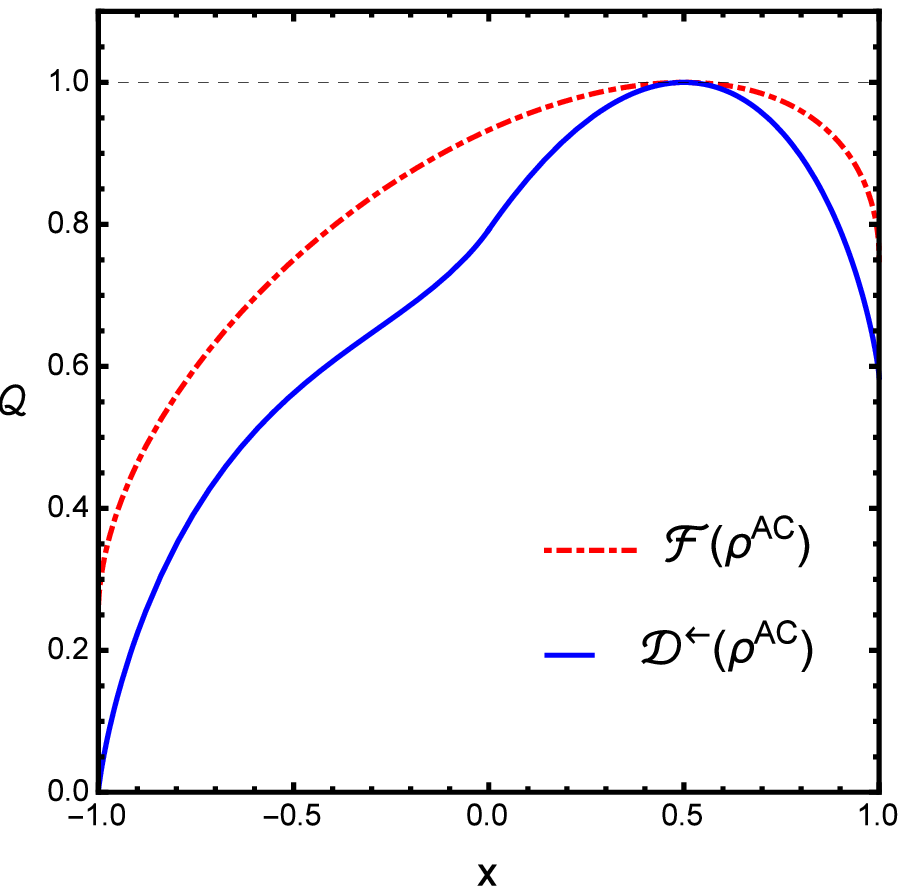}\label{22werner}}
\hspace{25mm}
\subfigure[]{\includegraphics[scale=0.7]{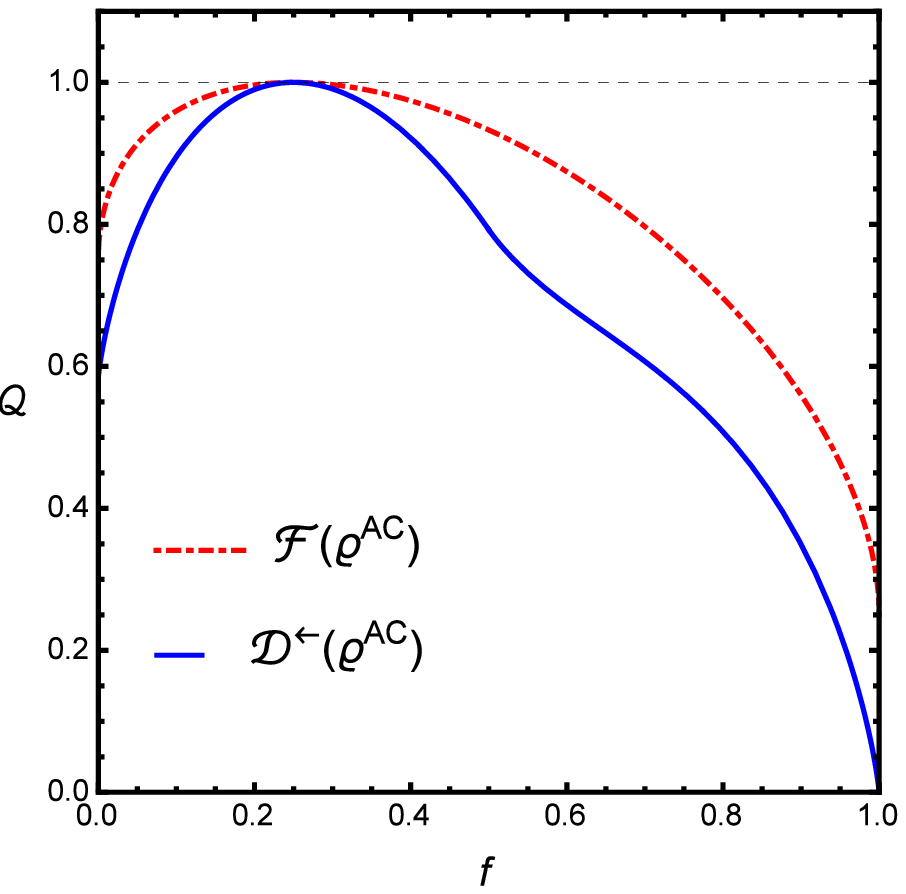}\label{22isotropic}}
\caption{(Color online) The fully entangled fraction and quantum discord of the $B$ complements to: \subref{22werner} $2\otimes2$ Werner states \subref{22isotropic} $2\otimes2$ isotropic states. The horizontal dashed line shows the maximal possible value $1$, both for the quantum discord and fully entangled fraction.}
\end{figure}

{\it   $B$ complement to  $2\otimes2$ Werner and isotropic states.---}
The $B$ complement  to   $2\otimes2$ Werner state is given by
\begin{equation}
\label{rhoac22w}
\rho^{AC}=\frac{1}{2}\ket{\Psi_1}\bra{\Psi_1}+\frac{1}{2}\ket{\Psi_2}\bra{\Psi_2},
\end{equation}
where
\begin{eqnarray}
\ket{\Psi_1}&=&\cos{\alpha}\ket{11}
+\sin{\alpha}\ket{2}\left[\cos{\theta}\ket{3}
-\sin{\theta}\ket{4}\right],\label{psi1} \\
\ket{\Psi_2}&=&\cos{\alpha}\ket{22}
+\sin{\alpha}\ket{1}\left[\cos{\theta}\ket{3}
+\sin{\theta}\ket{4}\right].\label{psi2}
\end{eqnarray}
In the standard basis, presented in Eq. \eqref{HCBasis}, on can write the $2\otimes 4$-dimensional state $\rho^{AC}$ as
\begin{equation}
\rho^{AC}=\begin{pmatrix}
\lambda_{+}&0&0&0&0&0&\frac{\lambda_{+}}{\sqrt{2}}&-\frac{\sqrt{\lambda_{+}\lambda_{-}}}{\sqrt{2}}\\
0&0&0&0&0&0&0&0\\
0&0&\frac{\lambda_{+}}{2}&\frac{\sqrt{\lambda_{+}\lambda_{-}}}{2}&0&\frac{\lambda_{+}}{\sqrt{2}}&0&0\\
0&0&\frac{\sqrt{\lambda_{+}\lambda_{-}}}{2}&\frac{\lambda_{-}}{2}&0&\frac{\sqrt{\lambda_{+}\lambda_{-}}}{\sqrt{2}}&0&0\\
0&0&0&0&0&0&0&0\\
0&0&\frac{\lambda_{+}}{\sqrt{2}}&\frac{\sqrt{\lambda_{+}\lambda_{-}}}{\sqrt{2}}&0&\lambda_{+}&0&0\\
\frac{\lambda_{+}}{\sqrt{2}}&0&0&0&0&0&\frac{\lambda_{+}}{2}&-\frac{\sqrt{\lambda_{+}\lambda_{-}}}{2}\\
-\frac{\sqrt{\lambda_{+}\lambda_{-}}}{\sqrt{2}}&0&0&0&0&0&-\frac{\sqrt{\lambda_{+}\lambda_{-}}}{2}&\frac{\lambda_{-}}{2}
\end{pmatrix}.
\end{equation}
Figure \ref{22werner} shows that the  fully entangled fraction of this state has the maximum  value $\mathcal{F}(\rho^{AC})=1$ at $x_{\max}=1/2$, which is the same point that quantum discord takes its  maximum value  $\mathcal{D}^\leftarrow(\rho^{AC})=1$.  In addition,  we observe two inflection points in the diagram of the quantum discord. One of the inflection points occurs in $x=0$ where is the separability-entanglement boundary  of the corresponding $2\otimes 2$ Werner state. In other words, this point is a witness for the boundary of separable states of the corresponding $C$ complement to $\rho^{AC}$, namely $\rho^{AB}_{\w}$. However, another inflection point can be observed in the diagram that we could not understand its physical justification.  Moreover, as it is clear from the figure \ref{22werner}, when the quantum discord vanishes, i.e. $x=-1$, fully entangled fraction takes its minimum value $\mathcal{F}=1/4$. As the minimum value of fully entangled fraction corresponds to the totally mixed state \cite{Zhao:2015}, we can infer that this class of states are not classical with respect the left subsystem, except for the case of $x=-1$, where the state  is classical with respect to the both subsystems. As it is known, the $2\otimes 2$ Werner and isotropic states are equivalent if we set $f=\frac{1-x}{2}$, so that  we expect their corresponding complemental states to show similar properties. Figure \ref{22isotropic} illustrates this resemblance.
Now let us consider the more complicated case of $3\otimes 9$  states.

\begin{figure}
\centering
\subfigure[]{\includegraphics[scale=0.7]{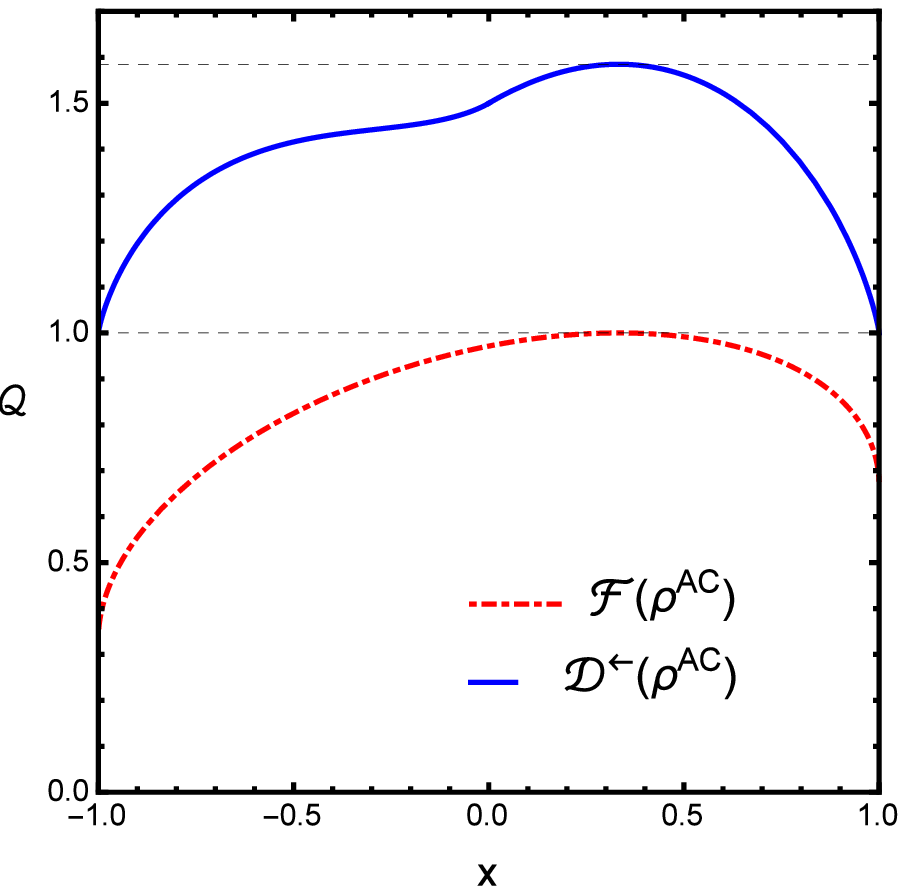}\label{33werner}}
\hspace{25mm}
\subfigure[]{\includegraphics[scale=0.7]{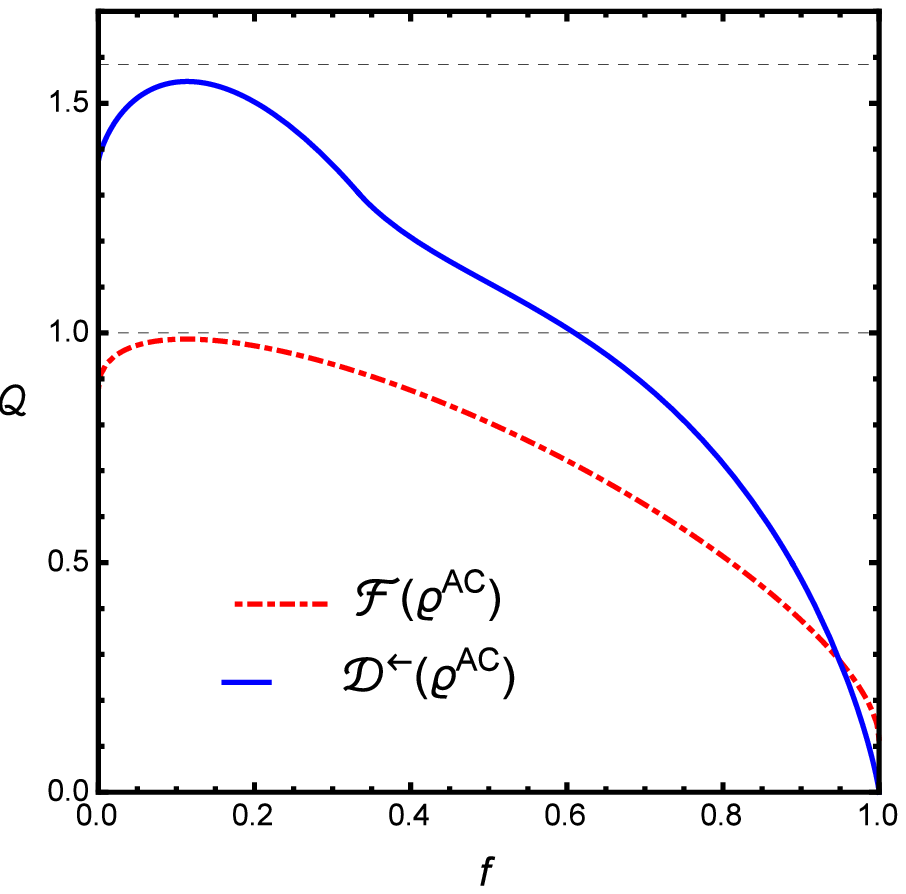}\label{33isotropic}}
\caption{(Color online) The fully entangled fraction and quantum discord of the $B$ complements to: \subref{22werner} $3\otimes3$ Werner states \subref{22isotropic} $3\otimes3$ isotropic states. The horizontal dashed lines at $\mathcal{Q}=1$ and $\mathcal{Q}=\log_2{3}$ show the maximal possible values for the fully entangled fraction and quantum discord, respectively.}
\end{figure}

{\it  $3\otimes 9$ complemental states of $3\otimes 3$ Werner states.---}
The complemental state corresponding to the $3\otimes 3$ Werner state is given by
\begin{equation}\label{Rho3x9W}
\rho^{AC}=\frac{1}{3}\ket{\Psi_1}\bra{\Psi_1}+\frac{1}{3}\ket{\Psi_2}\bra{\Psi_2}+\frac{1}{3}\ket{\Psi_3}\bra{\Psi_3},
\end{equation}
where
\begin{eqnarray}
\ket{\Psi_1}=\cos{\alpha}\ket{11}
&+&\frac{\sin{\alpha}}{\sqrt{2}}\ket{2}\left[\cos{\theta}\ket{4}
-\sin{\theta}\ket{7}\right]+\frac{\sin{\alpha}}{\sqrt{2}}\ket{3}\left[\cos{\theta}\ket{5}
-\sin{\theta}\ket{8}\right],
\end{eqnarray}
\begin{eqnarray}
\ket{\Psi_2}=\cos{\alpha}\ket{22}
&+&\frac{\sin{\alpha}}{\sqrt{2}}\ket{1}\left[\cos{\theta}\ket{4}
+\sin{\theta}\ket{7}\right]+\frac{\sin{\alpha}}{\sqrt{2}}\ket{3}\left[\cos{\theta}\ket{6}
-\sin{\theta}\ket{9}\right],
\end{eqnarray}
\begin{eqnarray}
\ket{\Psi_3}=\cos{\alpha}\ket{33}
&+&\frac{\sin{\alpha}}{\sqrt{2}}\ket{1}\left[\cos{\theta}\ket{5}
+\sin{\theta}\ket{8}\right]+\frac{\sin{\alpha}}{\sqrt{2}}\ket{2}\left[\cos{\theta}\ket{6}
+\sin{\theta}\ket{9}\right],
\end{eqnarray}
Figure \ref{33werner} illustrates the fully entangled fraction and quantum discord of the state \eqref{Rho3x9W}. As it is clear from this figure, the fully entangled fraction and quantum discord reach their corresponding maximum value at the same point, namely at $x_{\max}=1/3$. In addition, as we cannot find any states in this class with vanishing quantum discord, the fully entangled fraction never picks its minimal value, so that  $\mathcal{F}>1/9$. It is worth to mention that we can observe two inflection points, just like the case of $d=2$. Again, the one at $x=0$ reveals the separability boundary of the corresponding state complement to $\rho^{AC}$.

{\it  $3\otimes 9$ complemental states of $3\otimes 3$ isotropic  states.---}
The complemental state correspondig to a generic $3\otimes3$ isotropic state can be written as
\begin{equation}
\varrho^{AC}=p\ket{\Phi_1}\bra{\Phi_1}
+\frac{1-p}{2}\left(\ket{\Phi_2}\bra{\Phi_2}+\ket{\Phi_3}\bra{\Phi_3}\right); \qquad p=\frac{9-x}{24},
\end{equation}
where
\begin{eqnarray}
\ket{\Phi_1}&=&\cos{\beta}\ket{1}\left[\cos{\vartheta}\ket{\mu_{11}}-\frac{\sin{\vartheta}}{\sqrt{2}}\left(\ket{\mu_{22}}+\ket{\mu_{33}}\right)\right]
+\frac{\sin{\beta}}{\sqrt{2}}\left(\ket{2}\ket{\mu_{21}}+\ket{3}\ket{\mu_{31}}\right), \\
\ket{\Phi_{2}}&=&\cos{\tilde{\beta}}\ket{2}\left[\cos{\tilde{\vartheta}}\ket{\mu_{11}}+\sin{\tilde{\vartheta}}\ket{\mu_{22}}\right]
+\frac{\sin{\tilde{\beta}}}{\sqrt{2}}\left(\ket{1}\ket{\mu_{12}}+\ket{3}\ket{\mu_{32}}\right),\\
\ket{\Phi_{3}}&=&\cos{\tilde{\beta}}\ket{3}\left[\cos{\tilde{\vartheta}}\ket{\mu_{11}}+\sin{\tilde{\vartheta}}\ket{\mu_{33}}\right]
+\frac{\sin{\tilde{\beta}}}{\sqrt{2}}\left(\ket{1}\ket{\mu_{13}}+\ket{2}\ket{\mu_{23}}\right),
\end{eqnarray}
where $\beta$, $\tilde{\beta}$, $\vartheta$ and $\tilde{\vartheta}$ are defined according to the Eqs. \eqref{betatheta} and \eqref{betathetatilde}. It is clear that the fully entangled fraction for this class of states never reaches to its maximum value $1$, and according to the above discussion, we expect that the quantum discord never reaches to its maximum value $\log_2{3}$. Figure \ref{33isotropic} depicts fully entangled fraction and quantum discord of these states and confirms our guess. Again, just like the $B$ complement to Werner state, we can find two inflection points. The one at $f=1/3$ witnesses  the separability boundary  of the $C$ complement to $\varrho^{AC}$. In addition, at $f=1$, the state is totally mixed, so $\mathcal{F}=1/9$ and $\mathcal{D}=0$.

\section{CONCLUSION}
\label{s5}

We have investigated the quantum correlations of two classes of $d\otimes d^2$ states, complements to $d\otimes d$ Werner and isotropic states, via studying their fully entangled fraction and quantum discord. For the complement to $d\otimes d$ Werner state, we have found that the maximally entangled and discordant states are the same and occurred at $x_{\max}=1/d$. Moreover, the similar behaviour for the other values of $x$ confirms the compatibility of the predictions of fully entangled fraction and quantum discord. Despite the fact that neither the fully entangled fraction nor the quantum discord of state complement to $d\otimes d$ isotropic state becomes maximal, their maximum values occur at the same point, which supports again the compatibility of their behavior. The inflection points in the diagrams of quantum discord would be of physical importance. We showed that they can happen in the separability boundary  of the complements of the corresponding states. As a result, one would guess that the inflection points in the diagram of quantum discord can be exploited as a witness for the separability boundary of its complementary state.

\appendix
\section{Proof for lemma \ref{lem1}}\label{apa}
\begin{proof}We provide a proof for $2\otimes 4$ complemental state of a $2\otimes 2$ Werner state. The generalization to higher dimensions is straightforward. The states \eqref{psi1} and \eqref{psi2} can be rewritten as follows
\begin{eqnarray}
\ket{\Psi_1}=\cos\alpha\ket{11}+\sin\alpha\ket{2}\left(\cos\theta\ket{3}-\sin\theta\ket{4}\right),\\
\ket{\Psi_2}=\cos\alpha\ket{22}+\sin\alpha\ket{1}\left(\cos\theta\ket{3}+\sin\theta\ket{4}\right).
\end{eqnarray}
Having a glance of the above states, one can easily infer that the two maximally entangles states which maximize the fully entangled fraction of the state \eqref{rhoac22w} should be in the following form
\begin{eqnarray}
\ket{\psi_1^{\max}}=\frac{1}{\sqrt{2}}\ket{1}\left(\cos\gamma\ket{1}-\sin\gamma\ket{2}\right)+\frac{1}{\sqrt{2}}\ket{2}\left(\cos\gamma^\prime\ket{3}-\sin\gamma^\prime\ket{4}\right),\\
\ket{\psi_2^{\max}}=\frac{1}{\sqrt{2}}\ket{2}\left(\cos\gamma\ket{1}+\sin\gamma\ket{2}\right)+\frac{1}{\sqrt{2}}\ket{1}\left(\cos\gamma^\prime\ket{3}+\sin\gamma^\prime\ket{4}\right).
\end{eqnarray}
Now using the Eq. \eqref{FEFG} we have
\begin{equation}
\mathcal{F}=\max_{\{\gamma,\gamma^\prime\}}\left\{\bra{\psi_1^{\max}}\rho\ket{\psi_1^{\max}}+\bra{\psi_2^{\max}}\rho\ket{\psi_2^{\max}}\right\},
\end{equation}
which results to
\begin{equation}
\mathcal{F}=\frac{1}{4}\max_{\{\gamma,\gamma^\prime\}}\left(f_1+f_2\right),
\end{equation}
where
\begin{eqnarray}
f_1=\left(\cos\gamma\cos\alpha+\sin\alpha\cos\theta\cos\gamma^\prime+\sin\alpha\sin\theta\sin\gamma^\prime\right)^2,\\
f_2=\left(\cos\gamma\cos\alpha+\sin\alpha\cos\theta\sin\gamma^\prime+\sin\alpha\sin\theta\cos\gamma^\prime\right)^2.
\end{eqnarray}
It is easy to show that the maximum will be reached at $\gamma=0$ and $\gamma^\prime=\frac{\pi}{4}$ for an arbitrary $\alpha$ and $\theta$. The proof is completed and could be generalized to the higher dimensions in a straightforward procedure.
\end{proof}

\bibliography{BIB}{}

\end{document}